\documentstyle[11pt,hrd_pasp,twoside,epsf]{article}
\def\kelvin{{\rm \,K}}
\def\teff{\alwaysmath{T_{\rm eff}}}
\def\alwaysmath#1{\ifmmode{#1}\else{$#1$}\fi}
\def\feh{{\rm [Fe/H]}}
\def\mh{{\rm [M/H]}}

\def\etal{{et al.~}}

\def\ers{\alwaysmath{{\rm \, erg\,sec^{-1}}}}
\newcommand\msto{MSTO}

\newcommand\CHANDRA{{\sl CHANDRA~}}
\def\edcomment#1{\iffalse\marginpar{\raggedright\sl#1\/}\else\relax\fi}
\reversemarginpar

\begin{document}
\title{ Photometric properties of Stellar Populations in GGCs: a
multi-wavelength approach}
\author{ Francesco R. Ferraro}
\affil{Osservatorio Astronomico di Bologna, 
via Ranzani 1, 40127 Bologna, Italy}
 
\begin{abstract}
 Globular star clusters are extremely important astrophysical objects
since (1) they are prime laboratories for testing stellar evolution;
(2) they are ``fossils'' from the epoch of galaxy formation, and thus
important cosmological tools; (3) they serve as test particles for
studying the dynamics of the Galaxy; (4) information on individual
stars can provide constrains for stellar dynamic models; (5) they are
the largest aggregates in which all post Main Sequence (MS) stars can
be individually observed, and thus serve as fiducial templates for all
the studies of the integrated light from distant stellar systems.
In the framework of a long-term project devoted to 
 a multi-band study of stellar populations in Galactic Globular Clusters
  I present a set of recent results of a systematic study of
  (1) the Red Giant Branch in the near-IR (J,K)
      and in the optical bands;
  (2)  the Horizontal Branch  and Blue Stragglers stars
      from mid- and far-UV observations obtained with HST.

\end{abstract}

\section{Introduction}
 
Stellar evolution theory is crucial to yield a reliable clock
for dating astrophysical objects. Suitable  Color Magnitude Diagrams
(CMDs) and Luminosity Functions (LFs) are the most powerful tools to test
theoretical models and, in turn, the {\it running} of the stellar clock.
Within this framework, our group started a long--term project devoted to the
quantitative analysis and testing of each individual evolutionary sequence
in the CMDs of Galactic Globular Clusters (GGCs) in the most appropriate 
photometric bands (from the
near-IR to the far UV).  
The most recent results obtained  
are presented and discussed in this review.
In particular, I present results of a systematic study of
  {\it (i)} the Red Giant Branch (RGB) in the near-IR (J,K)
      and in the optical bands;
  {\it (ii)}  the Horizontal Branch (HB), Blue Stragglers stars (BSS)
  and some UV-excess stars (UVE) 
      from mid- and far-UV observations obtained with the {\it Hubble Space
      Telescope}.

\section{Cool Stellar Populations in GGCs}
 

The main aim of this part of the
 project is to obtain a complete quantitative description
of the RGB as a function of the intrinsic cluster parameters, and to
yield a few observational relationships for general use and suitable
to carefully test the theoretical models.
The full set of results were published in two recent papers:
Ferraro et al 1999  (F99) and Ferraro et al 2000a (F00).

\subsection{ The Optical catalog}

As a first step, we reviewed the papers on CMDs for GGCs
published over the last ten years, in order to select populous and
high quality CMDs. As a result, a catalog including  
the most recent  CMDs
 for a sample of 61 GGCs has been
  presented in F99.
We used this data-base to perform an homogeneous systematic
analysis of the evolved sequences (namely,
RGB,  HB and AGB).
 In F99, we presented:
(1) a new methodology to derive 
the actual level of the Zero Age Horizontal Branch (ZAHB) and 
the distance moduli from the matching
of V(ZAHB) and the theoretical models computed by Straniero, Chieffi 
\& Limongi (1997, hereafter SCL97);
(2) an independent estimate for RGB metallicity indicators
and  new calibrations of these parameters
in terms of both spectroscopic ($\feh_{\rm CG97}$) and
global metallicity ($\mh$, including 
also the $\alpha-$elements enhancement).
In particular,  we adopted the metallicity scale
presented by CG97  based on  high quality measurements of
iron abundances using high dispersion spectra of $FeI$ and $FeII$ lines.
However, in order to perform a correct
parametrization of the RGB behavior as a function of metallicity,
the simple knowledge of the quantity usually called {\it iron abundance}
is not sufficient. 
since the RGB location mainly depends on the [Mg+Si+Fe]
mixture abundance rather than on the [Fe] abundance alone. Therefore, a
more reliable parameter to describe the actual metal content of the RGB
stars is the so--called {\it global} metallicity, which takes into account
not only the iron but also the  
$\alpha$--element (like Mg and Si) abundance.
In doing this, we adopted the relation proposed by 
Salaris, Chieffi \& Straniero (1993).
  
\begin{figure}
\plotfiddle{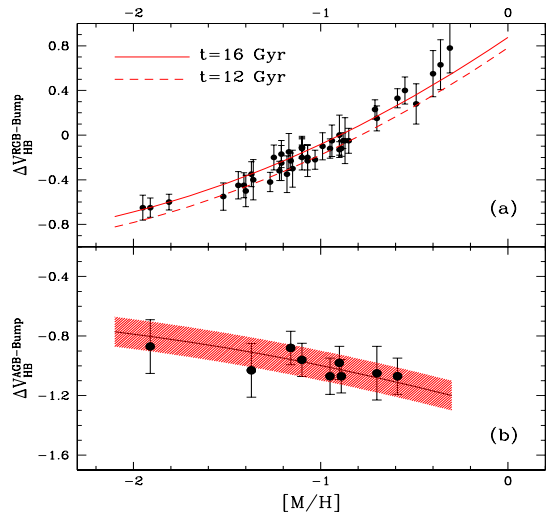}{6cm}{0}{42}{37}{-120}{-70}
\caption{ {\it Panel (a)}- The  
$\Delta V_{HB}^{Bump}$ parameter
  as a function of the global metallicity. The solid line
is the theoretical prediction
by SCL97 models at t=16 Gyr, the dashed line
represent the same set of models at t=12 Gyr.
{\it Panel (b)}-The difference between 
the observed ZAHB and AGB clump luminosity levels of
9 clusters. The solid line is the theoretical
expectation.
The shaded region is 
representative of the uncertainty ($\pm 0.1$ mag) in the absolute location
of the AGB-clump }
\end{figure}

In F99 we presented a complete set of equations which  
can be used to simultaneously derive
a {\it photometric} estimate of the metal abundance and the reddening
from the morphology and the location of the RGB in the $(V,B-V)$-CMD.
Moreover, we were able to determine the location of the  RGB-Bump
and the AGB-Bump in a number of GGCs.
In the following subsection I breafly summarize the 
results we got for these two
peculiar evolutive features along the Giant Branches.
    
\vskip3truemm
{\it The RGB-Bump --}The RGB 
evolution is characterized by a narrow burning hydrogen shell
which is moving towards the outer region of the star.
The shell is quite thin in mass and a temporary drop in luminosity is
expected when it reaches the discontinuity in the hydrogen distribution
profile generated by the inner penetration of the convective envelope.
This interruption in the expansion of the stellar envelope has its signature
in the differential LF star excess, the so--called bump.

Since the early work presented in Fusi Pecci et al 1990 (F90), 
the RGB-Bump has been
identified in a growing number of GGCs (see for example Brocato et
al. 1996).  As pointed out by  F90, the best tool to
identify the RGB-bump is the Luminosity Function (LF), and both the
integrated and the differential LFs are useful (Ferraro
1992). Following the prescriptions of F90 we independently identified
the RGB-bump in 47 GGCs.
{\it This is the
largest GGCs sample listing the RGB-bump locations available so far}.

To allow comparisons with   theoretical
models, following F90, we have measured the parameter: $\Delta
V_{\rm HB}^{\rm Bump}=V_{\rm Bump}-V_{\rm HB}$, which has the
advantage of being actually independent of the photometric zero-point
of the cluster data, the reddening, and the distance modulus.
In Figure 1(a) we compare the observed values of the 
$\Delta V_{\rm HB}^{\rm Bump}$ parameter with the latest
  theoretical predictions (SCL97)  
as a function of the global metallicity $[M/H]$.
 As a result, we found that by using the latest theoretical models
and the new metallicity scales the   discrepancy between 
theory and  observations ($\sim 0.4$ mag)
 found by F90, completely disappears.

\vskip3truemm
{\it The AGB-Bump --}
According to the evolutionary models (Castellani, Chieffi \& Pulone,
1991),   the beginning of the
AGB is characterized by a rapid increase of the luminosity 
followed by  a slowing down in the
evolutionary rate. Then, from an observational point of
view, a well defined clump of stars  is expected to
 indicate the base of the
AGB.

The models suggest 
(Castellani, Chieffi \& Pulone 1991, Pulone
1992) a luminosity level of the AGB-clump 
of $M_V^{\rm AGB-Bump}=-0.3\pm 0.1$, 
almost independent of the chemical composition of the cluster stars
(both $Z$ and $Y$), so that this (quite bright) feature could be a very
promising ``standard candle''.
  However, we note that the
theoretical calibration of the AGB clump location is affected by the
uncertainties in the actual extension of the convective core of an He
burning low mass stars. On the other hand, as pointed out by Caputo et
al. (1989) one might use the observed differences between the HB
luminosity level and that of the AGB-clump (i.e. $\Delta V_{\rm AGB}^{\rm
HB} =V^{\rm AGB}_{clump}-V_{\rm HB}$) to constrain the convection theory
(see
Dorman \& Rood 1993).
 
Unfortunately, the identification of such a clump is not easy since
the AGB phase itself is very short ($\sim 10^7$ yr) and, in turn,
always poorly populated.  There
are a few identifications of the AGB-clump in the literature: Ferraro
(1992) reported a preliminary identification of this feature in  a few
 GGCs. Then, to initiate 
 a systematic study of the properties of the AGB-clump,
in F99 we have identified such a feature in 9 GGCs whose CMDs
show a significant clump of stars in the AGB region.

\begin{figure}
\plotfiddle{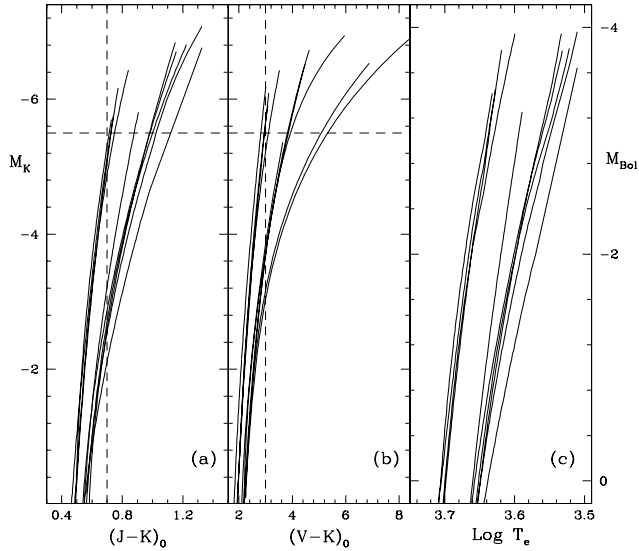}{7cm}{0}{46}{43}{-140}{-80}
\caption{RGB fiducial ridge lines for the 10 GGCs in the F00 sample in the
M$_K$,(J--K)$_0$,   
M$_K$,(V--K)$_0$
 and ($M_{Bol}, Log(T_{e})$)
  planes, ({\it panel (a)}, {\it  (b)} and {\it  (c)}), respectively.
The dashed lines indicate the magnitude levels at which some of the
parameters defined in  F99 are measured.
}
\end{figure}

In Figure 1(b), we compare the theoretical and observed values of the
$\Delta V_{\rm HB}^{\rm AGB}$ parameter. The shaded region represents
the quoted uncertainty ($\pm 0.1$) in the absolute location of the AGB
clump.  Despite the quite large error bar affecting most of the (few)
available measurements of the AGB-clump, the level of the agreement
with the theoretical prediction is remarkable.  Such a result, especially
combined with that obtained   for the RGB-bump location,
is comforting about the reliability and the internal consistency of
the adopted theoretical prescriptions.

\subsection{IR Observations}

The advantage of observing GGCs in the near IR is well known since many
years. 
The contrast between the red giants and the unresolved background
population in the IR bands is greater than in any optical region, so
they can be observed with the highest S/N ratio also in the innermost
region of the cluster. Moreover, when combined with optical observations,
IR magnitudes provide useful observables such as for example the
V--K color, which is an excellent indicator of the stellar effective
temperature (T$_{e}$) and allows a direct comparison with theoretical models
predictions.
 
 In F00 we presented 
 a new set of high quality IR Color Magnitude Diagrams   
  for a sample of 10 GGCs, spanning a wide range in metallicity.
This new, homogeneous data--base has been used to determine a variety of
observables quantitatively describing the main properties of the Red
Giant Branch, namely:
{\it (a)} the location of the RGB in the CMD (both in (J--K)$_0$ and 
(V--K)$_0$ colors at different absolute K magnitudes 
(--3, --4, --5, --5.5) and in temperature);
{\it (b)} its overall morphology and slope;
{\it (c)} the luminosity of the Bump and the Tip.
All these quantities have been measured via a homogeneous procedure
applied to each individual CMD.
Their behavior as a function of the cluster global  metallicity has been 
investigated.

\begin{figure}
\plotfiddle{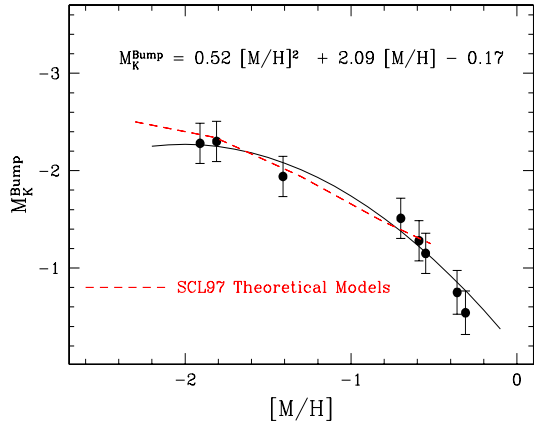}{5cm}{0}{40}{43}{-140}{-80}
\caption{Absolute K magnitude at the RGB bump as a function of
 the  {\it global}
metallicity scale,
for 8 GGCs in which the RGB--bump
has been identifiedby F00. The solid lines are the best fit to the data.
The dashed line is
the theoretical prediction by SCL97 models at $t=16$ Gyr. 
}
\end{figure}

In doing this we used
the distance moduli scale defined in F99  
to  locate the observed RGB fiducial ridge lines in the absolute
M$_K$,(J--K)$_0$ and M$_K$,(V--K)$_0$ planes. The result is  
plotted in Fig.2. The dashed lines in the figure represent
the magnitude levels and the colours at which some of the 
parameters defined in F00 are measured. The full discussion  of these
morphologic parameters and slopes 
and their  dependance from the metallicity
can be found in F00, here I will only 
show the results we got for the RGB-Bump
and the RGB-Tip (TRGB).
 
\vskip3truemm {\it The RGB-bump --} 
In Fig.3 the absolute K magnitude of the RGB-bump as a function of the
cluster global metallicity   $[M/H]$ 
is plotted. 
 The dashed line in  Figure 3  
   represents the theoretical expectations based on
the SCL97 models, for an age of $t=16$ Gyr (Straniero 1999, private
comunication). As can be seen, the models show an excellent agreement
with the observational data.
This result fully confirms the finding of F99
(from the location of the RGB bump in 47 GGCs in the visual band - see Fig. 1a)
that the earlier discrepancy between theory and observations ($\sim 0.4$mag)
(cf. F90, Ferraro 1992) has been completely removed
using the latest theoretical models and the {\it global} metallicity
($[M/H]$).

In order to allow a direct comparison  (and calibration) of the 
 theoretical models we transformed the fiducial lines plotted in Figure 2 in 
  the absolute $(M_{Bol},T_{e})$-plane by  using the bolometric
corrections and temperature scales for Population II giants
computed and adopted by Montegriffo et al. (1998) (cf. their Table 2).
In Fig. 2 ({\it  panel (c)}) the fiducial 
RGB ridge lines for the 10 GGCs in the F00 sample
are plotted in the M$_{Bol}$ {\it vs} T$_{e}$ theoretical plane.

\vskip3truemm {\it The RGB tip --}
A quite well defined relationship between the bolometric luminosity of
the brightest RGB star in a GGC and its metallicity has been found
by Frogel, Persson \& Cohen (1981) and Frogel, Cohen \&
Persson (1983, FCP83). This finding had a
noteworthy impact both on testing the theoretical models and on
the use of the brightest TGB stars as possible distance indicators.

In order to derive a similar relation using our sample of GGCs,
we identified the {\it candidate} brightest giant in each cluster, 
paying particular care in the decontamination of possible field
objects and (especially in the case of metal--rich clusters) of bright
AGB stars and the variables commonly associated to the AGB, the so--called
Long Period Variables (LPV). Of course, the degree of reliability of
the decontamination is hard to quantify due to the
very small number of stars populating the brightest extreme of the
giant branch, possibly affected by severe statistical fluctuations.
 
Fig.4 reports the bolometric magnitude of the  adopted
brightest star (the {\it observed} RGB tip)
 for 9 of the 10 clusters
considered in  F00 (M4 was excluded since the number of sampled
giants is too low). 

\begin{figure}
\plotfiddle{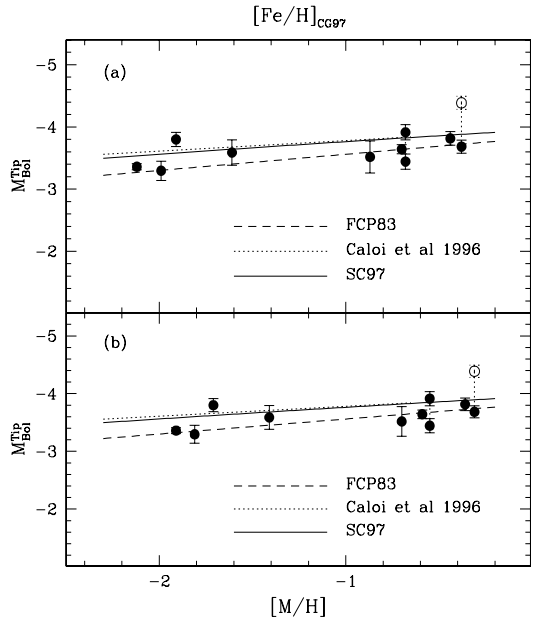}{7cm}{0}{40}{42}{-140}{-80}
\caption{ M$_{Bol}$  of the RGB tip as a function of
metallicity (in the CG97 and {\it global}
scale -- {\it panel(a) and (b)}, respectively)
for 9 GGCs discussed in F00.
The {\it dashed line} is the relation by FPC83. 
Two theoretical relations 
have been also plotted: Caloi et al. (1997)[{\it dotted line}] 
and Salaris \& Cassisi (1997, SC97) [{\it solid line}]. 
}
\end{figure}

As can be seen, our result is fully consistent with FPC83   
 (plotted as {\it dashed line} in Figure 4),
and with theoretical models. 
Two theoretical relations have
been over--plotted in Fig.4: Caloi et al. (1997) (dotted line)
and Salaris  \& Cassisi (1997, hereafter SC97) (solid line), respectively.
Indeed, the theoretical prediction nicely agrees with the
observations and, though residual contamination and statistical fluctuations
could still affect the sample,  the success
of the theory in reproducing the data seems quite rewarding. 
It may also be interesting to note
that such an agreement indirectly implies that the adopted
distances and reddening should not be affected by large errors.
It is important to remind here that the theoretical
relationships have to be considered indeed as upper limits to the
luminosity of the observed giants, because of the statistical fluctuations
affecting these poor samples (Castellani, Degl'Innocenti \& Luridana, 1993).

\subsection{The Global Test: the LF of M3}
 
One of the classical tests of stellar evolutionary calculations is
through the comparison of theoretical and observed stellar 
LFs. The LF for
the stars below the main sequence turn-off (\msto) can be related to
the stellar initial mass function and be used as a probe of stellar
dynamics within a cluster. The LF of stars after the \msto\  depends
primarily on the rate of evolution and provides a direct and
straight-forward test of evolutionary calculations.  
Features in the observed LF can
be related to interior structure, as for example the RGB-Bump 
(see Section 2.1).
Beside the bump, more  
 subtle observables are also present: the
slope of the LF below and above the LF bump are different because the
H-burning shell is in the first case passing through a region of
varying H abundance and later through a region of constant H
(F90,F99,F00). 
In particular,
a breakdown in {\em canonical} stellar
evolution theory (Renzini \& Fusi Pecci, 1988) can affect the
LF in the region of the subgiant branch (SGB).
Some preliminary evidences of the existence of  
 a breakdown in the
canonical models for very low metallicity clusters  was 
showed by Stetson (1991) and
Bolte (1994) and recently confirmed for M30 
(Sandquist \etal\ 1999).  This region could be affected by
non-canonical assumptions like WIMP energy transport (Faulkner \& Swenson 1993),
 and helium settling (SCL97), both of which
could reduce cluster age estimates. If large enough, also 
stellar rotation could
affect the region of the LF (VandenBerg, Larsen, \& de Propris 1998). 
  
In this scenario, a few years ago, we 
 started an ambitious project aimed to a new global approach to
the test of theoretical sequences: the immediate objective of this project was
the construction of a {\it new generation of LFs }  for a
selected sample of GGCs,
 in which {\it all post--MS stars at all radii}
 have been measured. The validity of this approach is shown by our work in
 the GGC M3. In this intermediate-metallicity cluster, we have constructed
 (by using a combination of ground-based
and HST observations) the   most complete  CMD 
ever obtained in a GGC.
 
\begin{figure}
\plotfiddle{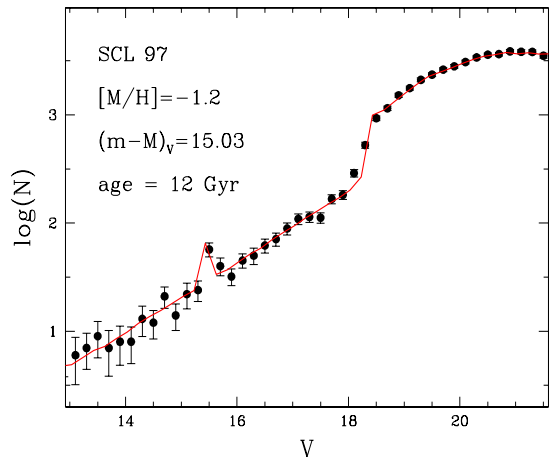}{5cm}{0}{40}{43}{-140}{-90}
\caption{ RGB differential luminosity function of the global sample
collected in M3
compared with theoretical models from SCL97. 
The chosen values for distance modulus, age and
chemical composition are indicated.  
}
\end{figure}

The data-base for this cluster  was collected
 during the last decades 
 (Buonanno et al 1994, Ferraro et al 1993, 1997a, 1997c)
 and the global LF constructed over this huge database 
 was finally recently published in Rood et al (1999). 
 The differential  LF for  the global sample is shown
in  Figure 5. {\it  Even
though it extends only 2 mag below the turnoff, this
LF, including more than 50,000 stars, is the most populated LF ever
published for a GGC}.
The solid line in the figure represent the SCL97 
theoretical  models for $(m-M)_V=15.03$, $\rm age = 12\,Gyr$  and  $ 
[M/H] = -1.2$. 
As shown in the Figure, the  theoretical and
observed LFs agree well in the SGB region and lower-RGB. Indeed the
fit  is essentially perfect: basically, there is no indication for
a breakdown in the canonical models.

\section{Hot Stellar Populations in GGCs}

As discussed in previous Sections, the CMD of a GGC
 in the {\it classical} $(V,B-V)$-plane is dominated by the cool
  stellar component. However, relatively populous hot stellar components 
  do
  exist in  GGCs
 and are significant emitters in the UV, 
namely: 
{\it (i)} the hot post AGB (PAGB) stars, 
{\it (ii)} the blue part of the HB, 
{\it (iii)} the BSS and, 
{\it (iv)}  various by-products of binary system evolution, such as
  interacting binaries (IB), for example Cataclysmic Variables (CVs),
Low Mass X-ray Sources (LMXB), etc. 
Some of these sources  (such as CVs, WDs, BSS, X-ray
sources, binaries, as well as the  complete luminosity function of the HBs) 
look hopelessly faint in the classical plane and for this
reason  remained unobserved until the advent
of the {\it Hubble Space Telescope}, whose space resolution and 
imaging/spectroscopic capabilities in the UV gave a new impulse to
these studies.

As a part of the global multi-band approach to the study of 
the stellar population in GGCs we are involved in a 
 long-term observational programme which uses
 the {\it  Hubble Space Telescope} to perform UV observations
 for a selected sample of GGCs.
 In this section I summarize   the most recent results about 
the UV populations (namely, HB, BSS, and faint UV sources).
  The advantages 
of studying  hot populations in the UV can be easily deduced from the 
comparison of  the two panels in Figure 1 of Ferraro, Cacciari
\& Paltrinieri (1999).

\begin{figure}
\plotfiddle{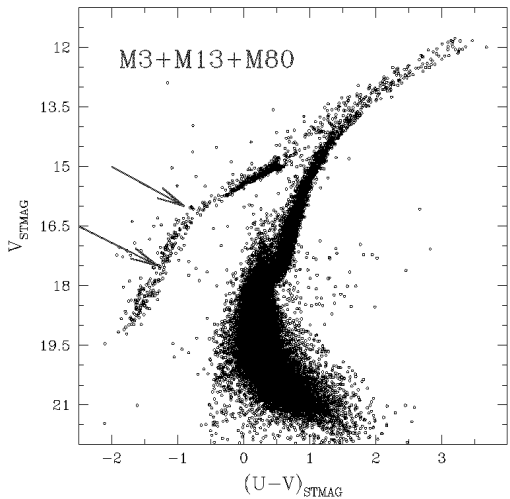}{7cm}{0}{40}{43}{-140}{-80}
\caption{The HB in the $(V,U-V)$ plane, for M3, M13 and M80, respectively,
after the alignment to M13. Only non-variable stars
have been plotted for M3.
Gaps along the HB are shown.
Note that in this plane the RR Lyrae
gap occurs at $(U-V)=0.55$ and it is not visible in the CMD
since it spans a small range in color. }
\end{figure}

\subsection{Gaps Along the HB: UV CMD to Proof Their Reality} 

Besides the global HB morphology difference between clusters at the same 
metallicity, another 
peculiarity was noted since the first CMDs of NGC 6752 (Cannon, 1981):
the discontinuous stellar distribution along the branch. The blue extension 
of the HB is, in fact,  often interruped by underpopulated regions or gaps.

The most recent evidence has been found in NGC 2808 (Sosin et al. 1997),
M13 (Ferraro et al 1997b) and M80 (Ferraro et al 1998). In particular,
the comparison of the CMDs of M13 and M80 showed 
that the HB in this two clusters do have similar characteristics:
 (1) the very long blue
tail extending $\sim 4.5$ mag 
(2) the non-uniform  stellar distribution, 
with at least 4 groups of stars
separated by gaps.  
 
 To get a better impression of the similarities in the overall HB morphology 
  we aligned the CMDs  in the $(V,~U-V)_{\rm STMAG}$ plane,
shifting the CMDs to match the M13 principal sequences and then
co-adding the result. We included also M3 in the comparison.
Figure 6 
shows the combined CMD, in which more than 50,000 stars have
been plotted. As can be seen from the small scatter along the RGB and
HB the sequences match well, showing the high degree of similarity of
the main branches in the CMD for these clusters:
the HB of M3
nicely matches the reddest part of the HB in M13 and M80.
Moreover, the HB
multi-group distribution is still (perhaps more obviously) present.
{\it This is the cleanest demonstration  that
the gaps observed along the HB are indeed real, since they
    have been observed at the same  location in different GGCs.}

A preliminary comparison with the theoretical models (Dorman \etal\, 1995)
and with other clusters (see Table 2 in Ferraro et al 1998) suggests that:

(a)  all the clusters with extreme long blue tail
have a gap on the lower blue tail 
at $\teff \sim 18,000\kelvin$. It  is present in at least 6 clusters
(M13, M80, NGC 6752, NGC 2808, $\omega$ Cen); 

(b) others gaps occur in many but not all clusters:
 the gap at  $\teff \sim 9,000\kelvin$ 
is present in 5 clusters (M13, M80, NGC6681, M79, M15), 
the gap at $\teff \sim 11,000\kelvin$  
is present in 3 clusters (M13, M80, NGC6681). 

Which is the scenario emerging from this picture?
The hottest gap occurs at a \teff\ at which the 
HB/post-HB evolutionary tracks change morphology:
the hotter stars (so-called extreme-HB or EHB) evolve
essentially vertically in the H-R diagram and do not return to the
asymptotic giant branch (AGB) after core He exhaustion
(AGB-manqu\'e stars). For this reason it 
is tempting to identify it  with the onset of EHB behavior
(Newell 1973, Rood 1997). 
No simple explanation can be  found for the  existence of the other
 gaps.
 
Evolutionary as well as statistical considerations seem to suggest that these 
gaps are produced by random mass loss efficiency during the RGB phase, which 
drives the mass distribution on the HB. Multiple mass loss mechanisms, or the 
multimodal behavior of a single mechanism, may be required in order to 
account for all the observed gaps
(see Ferraro et al 1998 for a full discussion of the problem).

\subsection{Blue Straggler Stars} 

Blue Straggler stars (BSS), first discovered by Sandage (1953) in M3, are 
interpreted as Main Sequence stars of larger mass, possibly originated by 
coalescence of binary/multiple stars, either from original systems or via 
later stellar collisions. Therefore they represent an important tool to 
study dynamical interactions inside star clusters. 
See Bailyn (1995) for a recent review on this subject.

In this context, the first surprising result
 has been obtained in the GGC M3. By using HST data and 
    complementary ground-based observations, we were able {\it for the first 
    time} to follow the BSS radial distribution over the entire  extension 
    of the cluster: it turns out to be {\it  clearly bimodal}
    (Ferraro et al 1993, 1997c).
    Moreover the  LF  for BSS in the inner region has been found to be
    significantly different from that  obtained  in the 
    external regions, suggesting that BSS in different environments could 
    have different origin: while BSS in the core could be originated
    by stellar interactions, BSS in the external regions could be
    a merging of primordial binary systems.

\begin{figure}
\plotfiddle{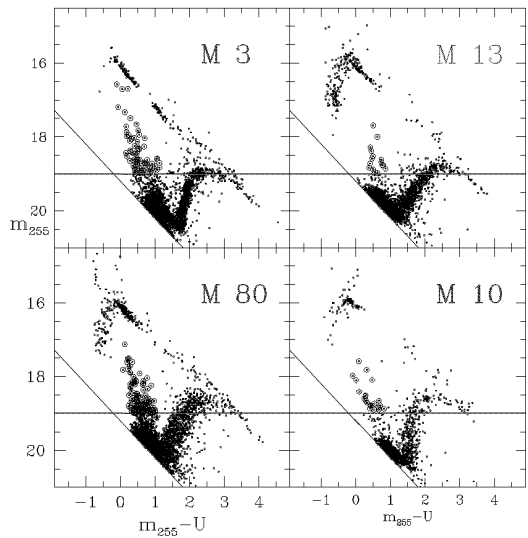}{7cm}{0}{40}{43}{-140}{-80}
\caption{  
($m_{255}, m_{255}-U$) CMDs for M3, M80, M92 and M13.
Horizontal and vertical shifts have been applied to M13, M80, and M92
to match the principal sequences of M3.
The horizontal  line corresponds to $m_{255}=19$.
The bright BSS candidates are plotted as large empty circles.
 }
\end{figure}

Encouraged by the results obtained in M3 we used our UV data-base to extend
the search for BSS in the core of three other cluster: M80, M13 and M10.
Figure 7 shows the ($m_{255},~m_{255}-U$) CMDs for the  four clusters:
more than 50,000 stars are plotted in four panels of Figure 7.  
Horizontal and vertical shifts have been applied to M13, M80 and M10
principal sequences in order to match those in M13.
The bright BSS $(m_{255}<19)$ are put in evidence in the Figure 
as big filled circles.
The BSS population selected in such a way (bright BSS) has been
compared to reference stellar populations (RGB and HB) observed 
in the same area of the cluster. The main results can be summarized as follows:
(1) BSS are more centrally concentrated than the corresponding RGB stars
in M3, M80 and M10, whereas in M13 
no significant difference has been found in 
the radial distribution of these stars;
(2) the BSS Specific Frequency (defined as $F_{BSS}=N_{BSS}/N_{HB}$)
turns out to be largely variable in the four clusters, varying by a factor 10
from M13 (the lowest) to M80 (the highest).
Particularly interesting is the case of M80
which shows an execeptionally high  BSS frequency: indeed, considering
the whole BSS content -- not only the bright BSS -- in the field of view of 
the PC, BSS turn out to be almost twice as abundant as HB stars. Indeed
BSS are overabundant in M80:
more than 300 BSS have been identified in this cluster (Ferraro et al 1999b): 
this is {\it the largest and most 
   concentrated BSS population ever found in a GGC}. 
   Since M80 is the  GGC which has the largest central density among those 
   not yet core-collapsed, this discovery could be the first direct evidence 
   that stellar collisions could indeed be effective in delaying the core 
   collapse.

These results seem to generally confirm the qualitative scenario proposed by 
  Ferraro, Bellazzini \& Fusi Pecci (1995) 
concerning the possible origin of BSS in globular 
clusters: the most important formation mechanisms would be coalescence 
of primordial 
binaries, most efficient in low density environments, and stellar collisions, 
most efficient in high density environments. Both mechanisms could be at 
work in different areas of the same cluster (e.g. M3), 
while the mechanism producing BSS via the collisional channel
should be at its maximum efficiency in M80 (Ferraro et al 1999b).

\subsection{Faint UVE} 

GGC cores have long been thought to harbor 
 other exotic objects which are 
 supposed to result from various kinds of binary systems:
low mass X-ray binaries, cataclysmic
variables, millisecond pulsars, etc. 
Though the number of such peculiar objects is admittedly low, they are
fundamental probes of GGC cores.  The systematic study of these
objects will permit a better understanding of the impact of
environmental conditions on stellar evolution.
In particular, some of them are most 
often connected to binary systems containing
compact objects (like neutron stars or white dwarfs) and are detected 
via X-ray emission. 
 
\begin{figure}
\plotfiddle{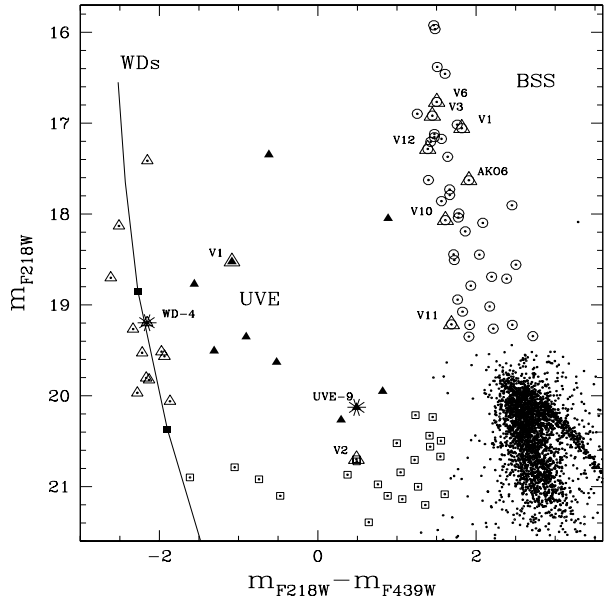}{7cm}{0}{45}{42}{-140}{-70}
\caption{  
 Zoom   of  the UV-CMD  of
 the  core of 47 Tuc. BSSs   are highlighted
by  large open circles,  with variables  further highlighted  by large
empty triangles  and labelled with  their names.  WDs  candidates  
   are plotted as  open triangles.  The theoretical  WD cooling
sequence from Wood (1995) models  has been overplotted to the data,
 with the  location of  3 and  13 million
year old cooling  WDs  are marked by filled  squares.  Bright UVE stars
  are plotted as filled triangles. The presence of a large
population of {\it  faint} UVE stars (empty squares)  is also shown.  
}
\end{figure}

In this respect, the faintest low luminosity X-ray sources (LLGCXs)
 ($L_X < 10^{32} \ers$) might
be associated with
cataclysmic variables (CVs), {i.e.,} binary systems in which a white dwarf
is accreting material from a late type dwarf,
for example a MS-SGB star (Hertz, Grindlay \& Bailyn, 
1993). 
The search for optical counterparts for LLGCXs is essential to
determine their origin and the role of the
dynamical history of the parent cluster.

Up to now, in all the  GGCs observed in our UV-HST survey 
we identified a few
faint objects with a strong UV excess standing significantly
outside the main loci defined by other cluster stars. Many of these
objects have been found to lie within the error boxes of LLXGCs
(see for example Ferraro et al 1997d, 2000c).
In a  recent review (Ferraro et al 2000b) 
 the photometric characteristics of  10 UVE objects,
associated to LLGCXs, detected in the
core of  6 GGCs have been compared and discussed.
Here I  report on  our most recent discovery:
  the surprising UVE population   in 47 Tuc.

In Ferraro et al 2001, we presented UV--CMDs
obtained for $\sim 4\,000$  stars detected within the Planetary Camera
(PC) field of view in the core of 47 Tuc.
We have pinpointed a number of interesting objects:
{\it (i)} 43 blue stragglers stars (BSSs) including 20 new candidates;
{\it (ii)} 12 bright (young) cooling white dwarfs (WDs) at the extreme
blue region of the UV-CMD; {\it (iii)} a large population of UV-excess
(UVE)  stars, lying  between the  BSS and  the WD  sequences.  
The
UVE  stars  discovered  in the core of 47 Tuc
   represent  the  largest  population  of
anomalous blue objects  ever observed in a globular  cluster -- if the
existence of  such a  large population is  confirmed, we  have finally
found the  long-searched population of  interacting binaries predicted
by the theory.  
A  zoom of  the ($m_{F218W},
m_{F218W}-m_{F439W}$) CMD is  shown in Figure 8: 
as it can
be seen   there are at least  a dozen of
UVE stars (plotted as filled triangles in Figure 8)
 lying between the WD and
the BSS  sequence and a large population of
faint UVE (small empty squares in Figure 8).
The  true   nature of these  {\it anomalous}
UVE  stars     cannot be 
satifactorily  assessed  yet.   In  fact many  different  evolutionary
mechanisms involving stellar collisions and interactions could account
for   them.  
High  resolution,   deeper  imaging   and  spectroscopic
observations in  the UV  are required to 
discriminate among the various models.
However, the large number of positional coincidences
with X-ray sources (from the recent \CHANDRA catalog by Grindlay et al. 2001)
 indeed suggests that part of 
them could be IBs. In particular, a significant number of X-ray sources
($\sim 30\%$) have been
found to be possibly associated with the {\it faint UVE} population
discovered here, supporting the presence of a large population of
faint CV in the core of this cluster.

\acknowledgments

It is a pleasure to thank  all the
   collaborators 
 involved in this  complex project, and E. Pancino  \& E. Sabbi for a 
 critical reading of
 the manuscript.
 The financial support of 
the Agenzia Spaziale Italiana (ASI) and  of the {\it Ministero della
Universit\`a e della Ricerca Scientifica e Tecnologica} (MURST) to the
project {\it Stellar Dynamics and Stellar Evolution in Globular
Clusters} is kindly acknowledged.

\end{document}